\documentclass[sigconf]{acmart}
\definecolor{red}{rgb}{0,0,0}
\newcommand{\clabel}[1]{\colorbox{black}{\textcolor{white}{\textsf{\textbf{#1}}}}}
\usepackage{xspace}
\usepackage{enumitem}
\usepackage{multirow}
\usepackage{fancyvrb}
\newcommand{\red}[1]{{\color{red}{#1}}}

\newcommand{\sys}{\textsf{AgentLens}\xspace}

\copyrightyear{2026}
\acmYear{2026}
\setcopyright{cc}
\setcctype{by-nc-nd}
\acmConference[UIST '26]{The 39th Annual ACM Symposium on User Interface Software and Technology}{November 02--05, 2026}{Detroit, MI, USA}
\acmBooktitle{The 39th Annual ACM Symposium on User Interface Software and Technology (UIST '26), November 02--05, 2026, Detroit, MI, USA}
\acmDOI{10.1145/3830398.3830619}
\acmISBN{979-8-4007-2856-3/2026/11}

\begin{document}
\title{AgentLens: Adaptive Visual Modalities for Human--Agent Interaction in Mobile GUI Agents}
\author{Jeonghyeon Kim}
\affiliation{%
  \department{Department of Computer Science and Engineering}
  \institution{Sungkyunkwan University (SKKU)}
  \city{Suwon}
  \country{Republic of Korea}}
\email{jeonghyeon12@skku.edu}

\author{Byeongjun Joung}
\affiliation{%
  \department{Department of Computer Science and Engineering}
  \institution{Sungkyunkwan University (SKKU)}
  \city{Suwon}
  \country{Republic of Korea}}
\email{bjbj2580@skku.edu}

\author{Junwon Lee}
\affiliation{%
  \department{Department of Computer Science and Engineering}
  \institution{Sungkyunkwan University (SKKU)}
  \city{Suwon}
  \country{Republic of Korea}}
\email{wwwnsdnjs@skku.edu}

\author{Joohyung Lee}
\affiliation{%
  \department{Department of Computer Science and Engineering}
  \institution{Sungkyunkwan University (SKKU)}
  \city{Suwon}
  \country{Republic of Korea}}
\email{jhl72e@skku.edu}

\author{Taehoon Min}
\affiliation{%
  \department{Department of Computer Science and Engineering}
  \institution{Sungkyunkwan University (SKKU)}
  \city{Suwon}
  \country{Republic of Korea}}
\email{mth9428@skku.edu}

\author{Sunjae Lee}
\affiliation{%
  \department{Department of Computer Science and Engineering}
  \institution{Sungkyunkwan University (SKKU)}
  \city{Suwon}
  \country{Republic of Korea}}
\email{sunjae.lee@skku.edu}
\begin{teaserfigure}
  \centering
  \includegraphics[width=0.8\textwidth]{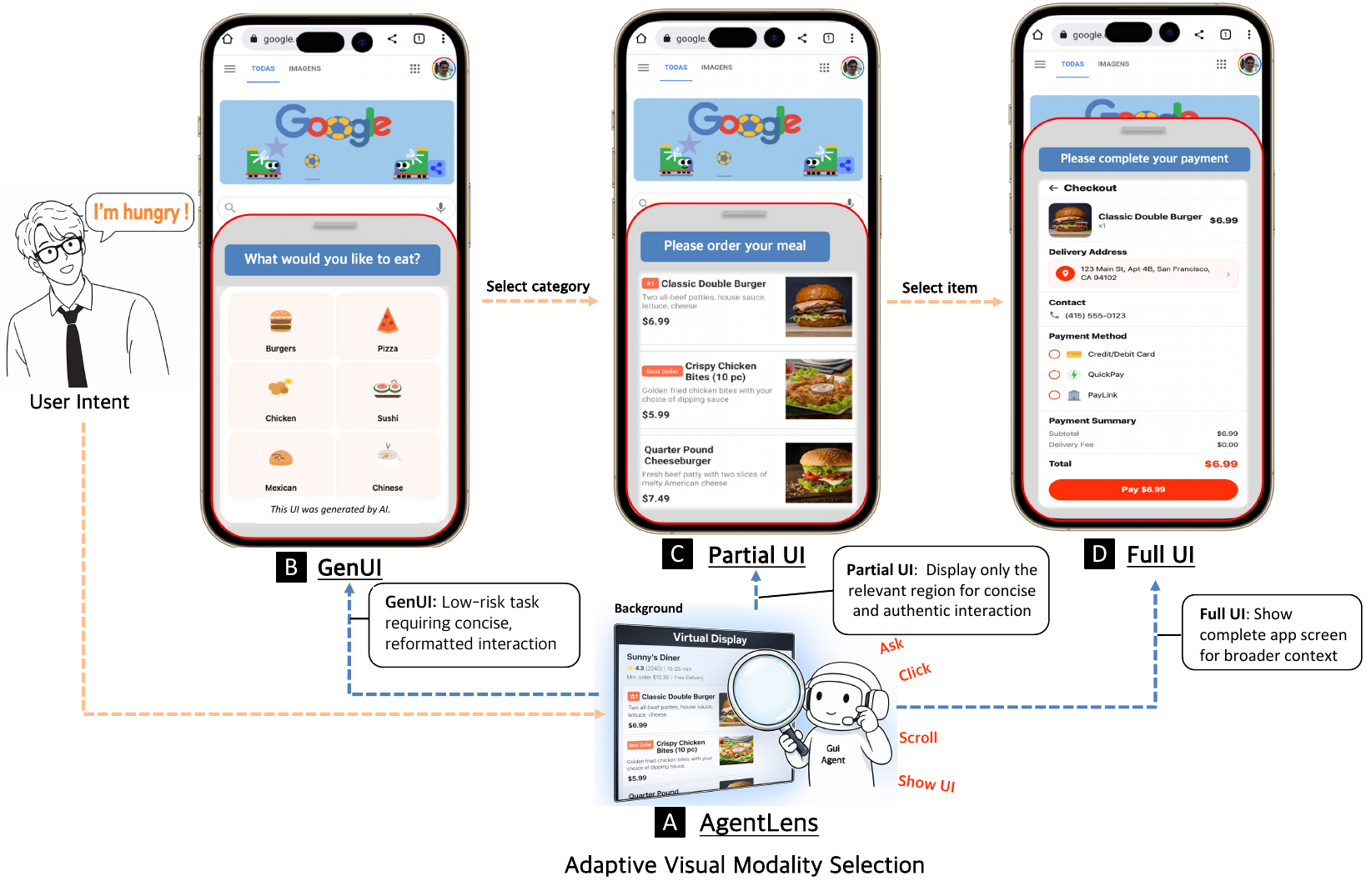}
  \vspace{-0.6cm}
  \caption{Overview of \sys{}. Given a user request (``I'm hungry!''), \sys{} \clabel{A} operates a delivery app in the background and adaptively selects among three visual modalities when user interaction is needed. \clabel{B} \emph{GenUI} presents an LLM-generated interface when a concise, reformatted interaction is most effective. \clabel{C} \emph{Partial UI} presents only the task-relevant region of the real app screen when authentic app content is needed but full-screen presentation would be unnecessarily intrusive. \clabel{D} \emph{Full UI} presents the complete original app screen when broader visual or spatial context is required, such as for verifying high-stakes actions at checkout. All overlays appear non-invasively over the user's ongoing activity (e.g., web browsing).}
  \Description{Overview of AgentLens with three phone mockups. Given the request "I'm hungry!", the agent operates a delivery app invisibly in the background and surfaces small popup overlays over the user's web browsing: an AI-generated food-category card (GenUI), the cropped menu region of the real app (Partial UI), and the complete checkout screen for payment verification (Full UI).}
  \label{fig:teaser}
\end{teaserfigure}

\begin{abstract}
Mobile GUI agents can automate smartphone tasks by interacting directly with app interfaces, but how they should communicate with users during execution remains underexplored. Existing systems rely on two extremes: foreground execution, which maximizes transparency but prevents multitasking, and background execution, which supports multitasking but provides little visual awareness. Through iterative formative studies, we found that users prefer a hybrid model with minimally invasive, just-in-time visual interaction, but the most effective visualization modality depends on the task. Motivated by this, we present \sys{}, a mobile GUI agent that adaptively uses three visual modalities during human--agent interaction: Full UI, Partial UI, and GenUI. \sys{} extends a standard mobile agent with adaptive communication actions and uses Virtual Display to enable background execution with selective visual overlays. In a controlled study with 21 participants, \sys{} was preferred by 85.7\% of participants and achieved the highest usability (1.94 Overall PSSUQ; lower is better) and adoption intent (6.43/7).
\end{abstract}


\renewcommand{\shortauthors}{Kim et al.}
\maketitle
\section{Introduction}
\label{sec:intro}
Mobile GUI agents---AI agents that operate smartphone applications on behalf of users by interacting with graphical user interfaces themselves---are rapidly emerging as a new mobile interaction paradigm. By perceiving on-screen content and executing touch-based actions such as tapping, scrolling, and typing, these agents can automate tasks across arbitrary mobile apps without requiring dedicated APIs or platform support. 

While mobile GUI agents hold the promise of transforming human-computer interaction, alleviating users from tedious and cumbersome smartphone operations,
recent progress in this area has focused primarily on the \textit{agent-side} of this pipeline---enhancing perception accuracy, augmenting reasoning capabilities, and boosting end-to-end task completion rates~\cite{mobile-agent-v2,mobilegpt,autodroid,verisafeagent,androidworld,appagent,guitars, coco-agent, mobileexperts, v-droid, android-inthe-zoo}. However, comparatively little attention has been paid to the \textit{user-side} of the interaction: \textit{how should a GUI agent communicate its progress, intentions, and queries to the user who delegated the task?}

Although a small body of HCI research has begun to explore \textit{when} agents should involve users---identifying points of intervention~\cite{Agent-initiated-interaction}, modeling delegation boundaries~\cite{delegation-boundaries}, and studying human--agent turn-taking---the question of \textit{how (i.e., with what modality}) the agent should present its intent to the user remains unexplored.

Today, mobile GUI agents typically operate in one of two extreme execution modes. They either run entirely in the \textit{foreground}~\cite{mobile-agent-v2}, occupying the full screen and visually exposing every step of their app interaction, or operate completely in the \textit{background}~\cite{apple-intelligence}, hiding the target app from the screen and communicating with the user solely through voice or text. The foreground execution mode provides full transparency---users can observe every action the agent takes---but it prevents multitasking and forces users to passively watch a slow, sequential automation process. The background execution mode, on the other hand, supports multitasking and parallel activity, but lacks visual feedback, leaving users unaware of what the agent is doing.

\textcolor{red}{One possible approach is to reuse the lightweight status widgets and rich notification interfaces that some mobile apps, such as food delivery or coffee-ordering services, provide for their background tasks. However, only a small subset of apps supports such interfaces, and each widget is inherently limited to a few fixed, developer-defined states and actions for a specific workflow. Extending this approach to mobile GUI agents is therefore neither broadly applicable nor scalable. Unlike app-specific background services, a mobile GUI agent is intended to operate across arbitrary apps and an open-ended action space that no finite collection of predefined widgets can anticipate. Supporting such agents instead requires interfaces that can either be generated on the fly or drawn directly from the underlying app to cover the full range of tasks and interactions that the app supports.}

One promising direction to address this gap is to leverage Generative UI (GenUI), a paradigm in which LLMs dynamically generate graphical user interface (GUI) to visually represent its intent to the user~\cite{genui-effective}. Applied to mobile GUI agents, GenUI could offer a middle ground between foreground and background execution modes by selectively generating and displaying a lightweight, non-invasive overlay UI only at critical moments, without occupying the full screen.

However, our iterative Wizard-of-Oz formative studies revealed that while users found the GenUI-based hybrid model---background execution with minimally invasive visual overlays at critical decision points---satisfying, many expressed trust concerns about LLM-generated interfaces. In particular, they worried that GenUI could hallucinate critical information such as prices, payment details, or menu options, especially in high-stakes tasks. In addition, participants wanted the agent to adapt how it visualizes its intent based on the task context. These findings led us to a broader design insight: the key question is not simply \textit{whether} to provide visual feedback, but \textit{what kind of} visual representation is most appropriate for the task at hand.

Guided by these findings, we present \sys{}, a mobile GUI agent system that adaptively employs three complementary visual modalities---Full UI, Partial UI, and GenUI---when human--agent interaction is needed. Full UI presents the live original app screen when broad visual or spatial context is needed; Partial UI shows only the task-relevant region of the real app to preserve authenticity while reducing distraction; and GenUI provides a generated interface when concise, reformatted interaction is more effective.

Architecturally, \sys{} extends a conventional mobile GUI agent with an adaptive visual interaction layer. Specifically, we augment the agent's action space with user-facing \texttt{speak} and \texttt{ask} actions, each paired with four visualization options---voice only, full UI, partial UI, and generative UI. In addition, to enable background execution of unmodified third-party apps and partial UI extraction from their interfaces, \sys{} leverages Android's Virtual Display~\cite{virtualdisplay} abstraction to operate the target app on an invisible surface, while selectively mirroring its cropped interface regions onto \sys{} companion app's popup overlays.

We demonstrate the effectiveness of \sys{} by implementing its agent prototype on top of the M3A system~\cite{androidworld}, a simple yet effective mobile GUI agent powered by GPT-5.4~\cite{gpt-5.4}, and its companion app on the Android platform\footnote{The system is available at
\url{https://github.com/kim-jeong-hyeon/AgentLens}}. We evaluated \sys{} through a controlled user study with 21 participants, comparing it against the two existing execution extremes, Foreground and Background. The results show that \sys{} is highly usable and practical in the real world: 18 of 21 participants (85.7\%) selected \sys{} as their first choice for daily use, and \sys{} achieved a high overall usability score, with \textbf{1.94 overall PSSUQ score} (lower the better) and 6.43/7 \textbf{adoption-intent score.}

To the best of our knowledge, \sys{} is the first work to explore the design space of visual interaction modalities between users and mobile GUI agents, and to propose technical solutions for realizing this design in a real mobile environment. Specifically, this paper makes the following three key contributions to the HCI community:

\begin{enumerate}[leftmargin=*, topsep=0pt, partopsep=0pt]
    \item Through iterative formative studies, we empirically uncover the underexplored design space of how mobile GUI agents should visually interact with users during task execution.

    \item We derive three design principles for visual interaction in mobile GUI agents and instantiate them in \sys{}, a system that adaptively selects among Full UI, Partial UI, and GenUI to present task-critical information in a non-invasive, just-in-time manner.
    
    \item Through a user evaluation with 21 participants, we validate the effectiveness of adaptive visual feedback in mobile GUI agents and distill implications for designing future mobile human--AI interaction.
\end{enumerate}
\section{Background and Related Work}
\label{sec:Related Work}

\subsection{Mobile GUI Agents}
Mobile GUI Agents have evolved from simple rule-based automation to sophisticated LLM-powered systems capable of understanding natural language instructions and interacting with complex mobile interfaces. Modern GUI agents typically operate through a perception--reasoning--action loop that translates user instructions into executable UI actions (e.g., clicks, scrolls, text inputs) across multiple steps:

\begin{itemize}[leftmargin=*, topsep=0pt, partopsep=0pt]
\item \textbf{Perception:} The agent converts the current mobile screen into a structured representation that the underlying model can reason over. Depending on the system, this representation may be derived from visual markers~\cite{som}, OCR~\cite{omniparser}, accessibility metadata~\cite{a11y}, or combinations of these signals.

\item \textbf{Reasoning:} Given the user instruction, and the perceived current screen, the model predicts the next UI action to execute. This stage is typically prompt-based and often incorporates systematic reasoning strategies such as ReAct~\cite{react} or few-shot prompting~\cite{few-shot-learning} to improve action grounding and task completion.

\item \textbf{Action:} Finally, the inferred action is translated into an actual input event on the device. This is commonly implemented either through ADB commands~\cite{autodroid, androidworld, autodroid-v2} or through accessibility service events~\cite{mobilegpt, verisafeagent}.
\end{itemize}
\textcolor{red}{This perception-reasoning-action paradigm now underlies systems ranging from general task automation to specialized applications such as accessibility testing~\cite{taeb2024axnav}.}
While this line of work has made rapid progress in improving execution accuracy, it has focused primarily on how agents interact with the \textit{GUI} itself. By contrast, our work investigates how such agents should interact with the \textit{user} during execution, particularly when the agent needs to inform, query, or request confirmation from the user while operating in the background.

\subsection{Human--Agent Interaction in Mobile Environment}

Mobile GUI agents operate within a uniquely constrained environment that demands distinct design considerations for human--agent interaction~\cite{foreground-background, porous-interfaces}. Current approaches either occupy the full display in foreground execution mode~\cite{appagent, autodroid, mobile-agent-v2} or run invisibly in background execution mode without visual feedback~\cite{apple-intelligence}. Research on supervisory control has shown that both extremes impose cognitive costs. Foreground execution reduces users to passively watching each action~\cite{out-of-the-loop}, while background execution without awareness cues forces users to tolerate suboptimal behavior or revert to active checking~\cite{complacency-automation}.

\textcolor{red}{Recent HCI work has begun mapping the UX design space of computer-use agents~\cite{cheng2026mapping} and characterizing the real-world impacts of agent-driven UI actions~\cite{zhang2025interaction}. These studies motivate keeping users informed and returning control to them at consequential moments~\cite{morae}, but leave open how task-relevant state should be visually presented on a mobile display.}

Recent industry efforts, such as Google's Gemini assistant~\cite{google-gemini-assistant}, have attempted to address this issue by combining background execution with lightweight status notifications. However, whenever the assistant needs to interact with the user (e.g., asking the user for input or confirmation), it still falls back to a full-screen takeover, switching between two execution modes rather than bridging the gap between them.

\textcolor{red}{Building on these insights, we explore an interaction model that bridges the two modes: \sys{} runs in the background and surfaces minimally invasive overlays only at decision-critical moments, adapting each overlay's form to the task. To our knowledge, \sys{} is the first mobile GUI agent to adaptively select among multiple visual modalities during human--agent interaction.}


\subsection{Visual Modalities for Human--Agent Interaction}
In the broader domain of AI agent design, various approaches have been explored for the communication between agents and users. The most prevalent modality remains the text-based chat interface~\cite{chatgpt, manus, claude}. However, text-based interfaces face fundamental limitations when agents must convey complex action states or collect multiple user inputs~\cite{genui-effective}. To address these shortcomings, a growing body of work has turned to Generative UI (GenUI), which leverages LLM to dynamically produce tailored interface components at runtime~\cite{generative-interfaces, a2ui, vercel-ai-sdk}. GenUI has since been explored across diverse HCI contexts, from malleable task-oriented workspaces~\cite{jelly} to ephemeral scaffolding within coding workflows~\cite{biscuit}.

Despite this promise, applying GenUI to more privileged, high-stakes agents such as mobile GUI agents raises various concerns, as LLM-generated interfaces remain susceptible to hallucinations and inconsistent adherence to user instructions~\cite{llms-meet-ui, duetui, genui-study}. Our work mitigates these risks by retrieving and displaying UI elements directly from the underlying app, eliminating the need for the generation in the first place, and restricting GenUI to low-stakes tasks.


\subsection{Partial UI Migration}
The idea of selectively displaying or migrating partial UI elements from one application to another has been explored in prior interface system research. Early desktop and web systems demonstrated this idea by extracting and migrating atomic UI components outside their original application context~\cite{wincuts, ui-facades, c3w, fusion}, and subsequent work in mobile environments, such as FLUID~\cite{fluid, fluid-xp} and A-Mash~\cite{amash}, extended this paradigm to small-screen devices. A substantial body of empirical work further supports the benefits of this approach: studies on peripheral displays have shown that partial information satisfies most awareness needs without requiring a full context switch~\cite{pielot-notifications, notifications-awareness}; research on adaptive interfaces confirms that selectively surfacing relevant functionality improves user satisfaction~\cite{adaptive-ui-design-space, spatial-consistency}. In light of these findings, \sys{} leverages \textit{Partial UI} display as one of the three primary communication channels between the agent and the user.



\section{Formative Studies}
\textcolor{red}{
To investigate effective human--agent interaction models for mobile GUI agents, we conducted a series of iterative formative studies using a Wizard-of-Oz (WoZ) methodology. In particular, we explored the applicability of Generative UI (GenUI), an emerging interaction paradigm that can dynamically generate task-specific interfaces, while remaining open to alternative interaction models that might better support communication between users and agents. Specifically, our formative studies addressed the following research questions:
}
\begin{itemize}[leftmargin=*, topsep=0pt, partopsep=0pt]
\item \textbf{RQ1 (Formative Study 1):} What interaction model do users want from a mobile GUI agent?

\item \textbf{RQ2 (Formative Study 2):} \textcolor{red}{How do users respond to a GenUI-based interaction model?}

\item \textbf{RQ3 (Formative Study 3):} What visual presentation modality is most appropriate under different task conditions?
\end{itemize}

\textcolor{red}{\subsection{Study 1: Exploring User Interaction Needs}}
\subsubsection{Procedure.}
To understand how users prefer to interact with a mobile GUI agent while delegating a task, we recruited \textcolor{red}{10 participants (\(P1\)--\(P10\); 3 female, 7 male; aged 20--25) through an online university community for a within-subject WoZ study. All were daily smartphone users, and reported high familiarity with chat-based AI assistants ($M$ = 6.9/7). Each participant was compensated $\sim$ \$15 USD per hour.} The study involved two experimenters: one remained in the same room as the participant and conducted the interview, while the other stayed in a separate room and remotely controlled the participant's smartphone to simulate the agent's behavior.

\textcolor{red}{To explore how users viewed the two existing execution modes, participants experienced both in counterbalanced order within a food-delivery scenario: Foreground execution mode (FG), in which the agent occupied the full screen and exposed its entire manipulation process, and Background execution mode (BG), in which it operated invisibly. Participants interacted via voice only, and the WoZ operator introduced an approximately 20--30 second delay at each step transition to reflect the per-step inference latency reported for current LLM-based agents.
We conducted semi-structured interviews after each condition, and a final one after both concluding with an open-ended question: \textit{``What would your ideal agent look like?''}}

\subsubsection{Findings.}

\textcolor{red}{\textbf{F1-1. Foreground was trusted but seen as impractical for everyday use.}
All 10 participants reported trusting the agent in FG because they could visually verify its actions. Yet none described it as a mode they would adopt in everyday life. Nine found the full-screen takeover unacceptable because it blocked all other phone activity, and seven were frustrated at having to watch the agent's slow, step-by-step operation. As P10 put it, \textit{``I could just tap through it myself faster. Watching the agent go through each voice command, then process, then act... that loop takes too long.''} This points to a tension in FG-based GUI agents: full transparency comes at the cost of usability.
}

\textcolor{red}{\textbf{F1-2. Background execution mode supported multitasking but raised anxiety and cognitive load.} BG was widely preferred for its multitasking support (n = 9), but it also raised concerns. Nine participants worried about irreversible tasks such as payments, where voice confirmation alone felt insufficient; P9 stated, \textit{``Messing up an email is fine, but a wrong payment is very hard to undo---I need to at least \textbf{see} the final screen before it goes through.''} Six participants reported cognitive load from voice-only delivery: when the agent read out multiple options sequentially, they struggled to retain and compare them, with P4 and P5 noting, \textit{``When there are lots of options, I just can't remember them.''} This suggests that voice alone struggles when users need to compare options at once.}

\textcolor{red}{\textbf{F1-3. Participants independently gravitated toward a hybrid model.}
When asked to describe their ideal agent, without any alternative suggested by the researchers, all 10 participants consistently described a middle ground combining FG's visual reassurance with BG's multitasking convenience. Their responses clustered around three recurring themes: \textit{i) minimally invasive UI}: non-intrusive overlays such as small popups (P2, P3), a Dynamic-Island-style element (P4), or toast notifications (P10)---P2 asked for it \textit{``small, like a popup''}, and P10 wanted it \textit{``just slipped in, like a little toast''}; \textit{ii) selective visibility}: surfacing only the relevant information rather than the full app screen (P1, P3, P4, P5, P8, P10), as P8 put it, \textit{``the whole screen isn't necessary—just show the information I need''}; and \textit{iii) just-in-time intervention}: visual UI appearing only at moments requiring user input (P2, P3, P6, P7, P8, P10)---P3 wanted to \textit{``see only what's related to the payment, right when I pay.''}
}

\textcolor{red}{\subsection{Study 2: Evaluating User Experience with a GenUI Prototype}}
\subsubsection{Procedure.}
Immediately after Study 1, the same 10 participants experienced a WoZ prototype implementing our initial design hypothesis: Generative UI based mobile GUI agent. In this condition, the agent ran in the background by default, but whenever user involvement was needed, it generated and presented a lightweight GUI overlay accompanied by a short voice notification. The GUI overlay was pre-generated in HTML using GPT-5.4, following a common Generative UI workflow~\cite{genui-effective}. Participants could respond either by tapping the overlay or through voice.

After the session, we conducted semi-structured interviews evaluating the prototype's acceptability, and their perception of the generated interface. We then explicitly disclosed that the interface had been generated by an LLM, and probed how this knowledge affected their assessment.

\subsubsection{Findings.}

\textcolor{red}{\textbf{F2-1. The hybrid model was widely preferred.}
Nine of ten participants preferred the GenUI-based hybrid model over both FG and BG for daily use, describing it as combining BG's multitasking support with FG's visual reassurance at critical moments. P9 noted that the overlay \textit{``cut the long voice listings down to a quick glance and a tap.''} P1, the exception, still preferred BG, finding the overlay intrusive: \textit{``the UI kept coming up and getting in the way of what I was watching.''}
}

\textcolor{red}{\textbf{F2-2. GenUI raised hallucination concern.}
When we disclosed that the overlays were generated by an LLM, reactions diverged. Seven participants raised concerns. Three (P1, P4, P5) expressed strong distrust: \textit{``What if it makes up a menu item that doesn't exist? Then I'd be charged for the wrong thing''} (P5); \textit{``If I experienced a hallucination even once, I'd never trust it again''} (P4). Four (P2, P6, P7, P10) raised conditional concerns, finding GenUI acceptable for simple tasks but not for critical contexts such as finance. The remaining three (P3, P8, P9), who reported more technical background, were less concerned; P8 noted, \textit{``If it generates the UI from data extracted from the app, that should be reliable enough.''}
}

\textcolor{red}{\textbf{F2-3. GenUI has unique strengths, but the question of how to present it remains open.}
Participants recognized strengths unique to GenUI—surfacing details easy to miss and reshaping content that would not fit a small overlay. As P9 put it, \textit{``The real app was designed for a full screen; if you'd dropped it in as-is, I don't think I could have read it.''} Even so, when asked how the content should be presented, preferences did not converge. Four preferred the real app over a generated one, wary of hallucination; as P4 put it, \textit{``I'd trust the real app more—I don't get why it needs to be regenerated; couldn't it just show the whole app screen as a capture?''} Two preferred GenUI, two said it depended on the context, and the remaining two expressed no clear preference. These observations left us with an open question: in what way should the agent visually interact with the user? This question motivated our next study.
}

\subsection{Study 3: Comparing UI Modalities}
\subsubsection{Procedure.}
{\color{red}{Studies 1 and 2 suggested that users wanted a hybrid model---non-invasive visual feedback at decision-critical moments---but left open what to show and how. To explore this, we compared three visual modalities that had emerged from participants' descriptions across those studies: \textsf{Full UI}, mirroring the app's entire screen; \textsf{Partial UI}, showing only the task-relevant region of the real app; and \textsf{GenUI}, reconstructing the interface through LLM generation. The complete screenshots for all scenarios and modalities are in \autoref{Appendix:B}.

We recruited a new cohort of 10 participants (\(P11\)--\(P20\); 7 female, 3 male; aged 18--25) through the same online university community. All were daily smartphone users, and reported high familiarity with chat-based AI assistants ($M = 6.1/7$). Each participant was compensated $\sim$ \$15 USD per hour. They compared the three modalities across six scenarios---food delivery, music streaming, online shopping, weather, financial account inquiry, and taxi hailing---ranking the modalities and explaining their reasoning in semi-structured interviews. Scenario order and modality order were fully counterbalanced across participants.
}}
\subsubsection{Findings.}
\textcolor{red}{\textbf{F3-1. No single modality was universally preferred.}}
Across 60 total ratings (10 participants \(\times\) 6 scenarios), no single modality dominated. First-place preferences were distributed relatively evenly: \textsf{Full UI}, 31.7\% (19/60); \textsf{Partial UI}, 36.7\% (22/60); and \textsf{GenUI}, 31.7\% (19/60). These results suggest that there is no one-size-fits-all modality strategy.

\textcolor{red}{\textbf{F3-2. Task characteristics influence modality preference.}}
Preferences showed distinct patterns based on the scenario. \textsf{Full UI} was preferred when broad app context or spatial layout mattered, such as checking the weather or calling a taxi. P16 explained: \textit{``When I call a taxi, I need to check the map, price, and available options before making a decision.''} \textsf{Partial UI} was favored when users needed trustworthy, precise information but not the entire app. P12 noted: \textit{``I like how it shows only the necessary information directly from the app. I don't feel comfortable with AI regenerating my financial information.''} \textsf{GenUI} was preferred when the task was low-risk and the original app's UI was seen as suboptimal. P11 commented (under music app scenario): \textit{``I always thought that album covers are unnecessary. The clean list of titles is much better.''}

\textcolor{red}{
\textbf{F3-3. Participants wanted the agent to switch modalities adaptively.}
Most participants (n = 8) wanted the agent to switch presentation modality depending on the task rather than rely on a single fixed form: \textit{``I would want the agent to select the best modality for me''} (P17); \textit{``I hope the agent handles it on its own based on the task, preference, and context''} (P16).
}

\begin{figure*}[h!t]
    \centering
    \includegraphics[width=0.9\textwidth]{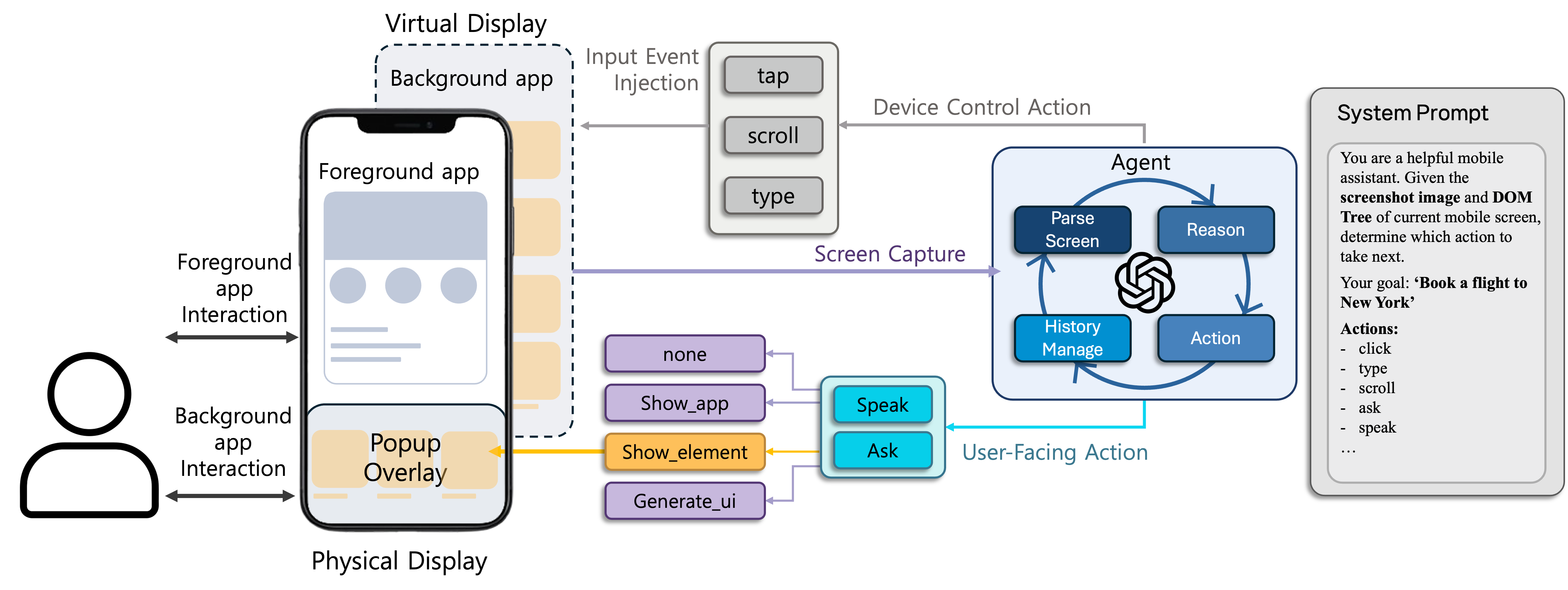}
    \vspace{-0.5cm}
    \caption{\sys{} System Architecture. \sys{} uses Virtual Display to operate third-party apps in the background, while displaying the visual feedback (i.e., overlay) on the physical display.}
    \vspace{-0.5cm}
    \label{fig:architecture}
    \Description{AgentLens system architecture diagram. The agent captures screens from a Virtual Display hosting the background app and injects taps, scrolls, and typing back into it, while speak and ask actions with one of four visualization options (none, show_app, show_element, generate_ui) are rendered as a popup overlay on the user's physical display.}
\end{figure*}

\section{\sys{}}
\label{sec:method}
Our formative studies revealed that users want a mobile GUI agent that operates in the background, surfaces visual information only at decision-critical moments, and adapts the form of that information to the task at hand. To the best of our knowledge, no existing mobile GUI agent supports such adaptive, multi-modal visual interaction with the user. In this section, we first present the design principles distilled from our formative findings, then describe the architecture of \sys{}.

\subsection{Design Principles}
\paragraph{DP1. Background execution with non-invasive visual intervention.}
The agent should run in the background by default. When visual information is needed, it should be presented through a lightweight overlay rather than a full-screen takeover (F1-1, F1-3, F2-1).

\vspace{-0.5\baselineskip}
\paragraph{DP2. Just-in-time intervention at decision-critical moments.}
Visual overlays should appear only when the user's input, confirmation, or awareness is required---such as when selecting among multiple options, approving an irreversible action, or checking a task result (F1-2, F1-3, F2-1).
\vspace{-0.5\baselineskip}
\paragraph{DP3. Adaptive selection of UI presentation modality.}
No single visual modality suits all tasks. The agent should dynamically select among Full UI, Partial UI, and GenUI based on the task context (F2-3, F3-1, F3-2, F3-3).

\subsection{Agent Overview}
Building on the standard pipeline architecture of mobile GUI agents introduced in ~\autoref{sec:Related Work}, \sys{} extends a conventional GUI agent with an adaptive visual interaction layer (see~\autoref{fig:architecture}). In particular, while the core \textit{perception--reasoning--action} loop remains unchanged, \sys{} augments the agent's action space with explicit mechanisms for interacting with the user during task execution.

Specifically, in addition to standard device-control actions such as taps, scrolls, and text input, \sys{} introduces two user-facing action types: \textsf{speak} and \textsf{ask}. The \textsf{speak} action is used when the agent needs to notify, summarize, or explain information to the user without requiring a response. The \textsf{ask} action is used when the agent requires user input, confirmation, or selection before it can proceed. Both \textsf{speak} and \textsf{ask} are associated with a \textit{visualization option} parameter that determines how the relevant information should be presented. The agent can choose among the following four visualization methods:
\begin{itemize}[leftmargin=*]
    \item \texttt{none}: The message is delivered through voice only.
    \item \texttt{show\_app}: Presents the full app screen to the user (\textbf{Full UI}).
    \item \texttt{show\_element}: Extracts and displays only the task-relevant region of the target app screen (\textbf{Partial UI}).
    \item \texttt{generate\_ui}: Presents an LLM-generated interface tailored to the current interaction need (\textbf{GenUI}).
\end{itemize}

When the agent outputs a \textsf{speak} or \textsf{ask} action with a visualization option other than \texttt{none}, the \sys{} companion mobile application renders the corresponding UI as a popup overlay, as shown in ~\autoref{fig:system_example}. We engineered the agent's prompt to reflect the design principles, so that the agent invokes these actions only at decision-critical moments and selects an appropriate visualization mode based on the findings from our formative study (see ~\autoref{Appendix:A} for the full prompt).

\subsection{\sys{} System Design}
Realizing the agent described above requires addressing two technical challenges. First, the system must be able to operate arbitrary third-party mobile apps in the background while keeping them fully functional. Second, it must extract and present only specific parts of an application's interface in order to support Partial UI.

\textbf{C1. Background App Execution via Virtual Display.}
A key design goal of \sys{} is to operate apps without taking over the user's screen. However, this is non-trivial in mobile operating systems, as applications moved to the background are typically deprioritized---stop rendering, become idle, or cease to receive input events---for resource efficiency.

The conventional approach to enable such non-invasive assistant interaction is to integrate a target app with a vendor-provided system-level assistant (e.g., Google Assistant, Siri, Bixby) through dedicated APIs. However, this requires modifying the app's source code, which contradicts the fundamental premise of GUI agents: automating any third-party app transparently. Worse yet, such integration is limited to a few pre-defined categories of apps such as messaging, alarms, or navigation~\cite{bixby_api, app_action}.

To address this challenge, \sys{} leverages the \textit{Virtual Display} abstraction that exists on mobile platforms under different names (e.g., VirtualDisplay~\cite{virtualdisplay}). A Virtual Display is a software-created display surface that behaves like a secondary screen that exists only within the system. For our purposes, they can be utilized as an invisible and isolated execution environment in which an app can be launched without interfering with the user's physical display.
Specifically, when \sys{} receives a request to launch a target app, rather than opening it on the physical display and taking over the screen, it creates a Virtual Display \textcolor{red}{through Android's \texttt{DisplayManager} ~\cite{displaymanager} API and launches the app onto it by issuing an activity-start command targeted at that display.} The agent's perception--action loop then operates over this virtualized environment: screenshots and \textcolor{red}{the accessibility tree} are captured from the Virtual Display, and input events (e.g., taps, scrolls) are injected by forwarding them to the Virtual Display \textcolor{red}{via ADB (Android Debug Bridge)}. This gives \sys{} a fully functional, isolated execution environment that operates entirely in the background (see~\autoref{fig:architecture}).

We note that this mechanism may appear to pose a security concern, as it could in principle allow a malicious agent to silently manipulate a user's apps. \red{In practice, however, launching a third-party app on a Virtual Display can only be done through ADB commands, which are accessible only when the device is explicitly connected to a desktop computer with developer options enabled.}
While our implementation is a research prototype, a production deployment could be realized through a system-privileged mobile assistant app with appropriate platform permissions.

\textbf{C2. Partial UI Visualization through Cropped Mirroring.}
To support the \textsf{Partial UI} modality, the system must be able to extract and present only a specific UI element from the target app. Although prior work on UI migration has explored splitting or relocating individual UI elements across devices or applications~\cite{fluid,fluid-xp,amash}, these approaches require either modifications to the app or to the Android operating system itself.

\sys{} addresses this by cropping the task-relevant region from the Virtual Display and mirroring it onto the overlay surface. To specify which region to display, \sys{} follows the common practice of feeding the LLM the app's accessibility node tree alongside the screenshot and indexing each element with a numeric identifier, so that the agent can specify UI elements by index rather than by pixel coordinates~\cite{androidworld, autodroid, mobibench}.

\begin{figure}[t]
    \centering
    \includegraphics[width=1.0\linewidth]{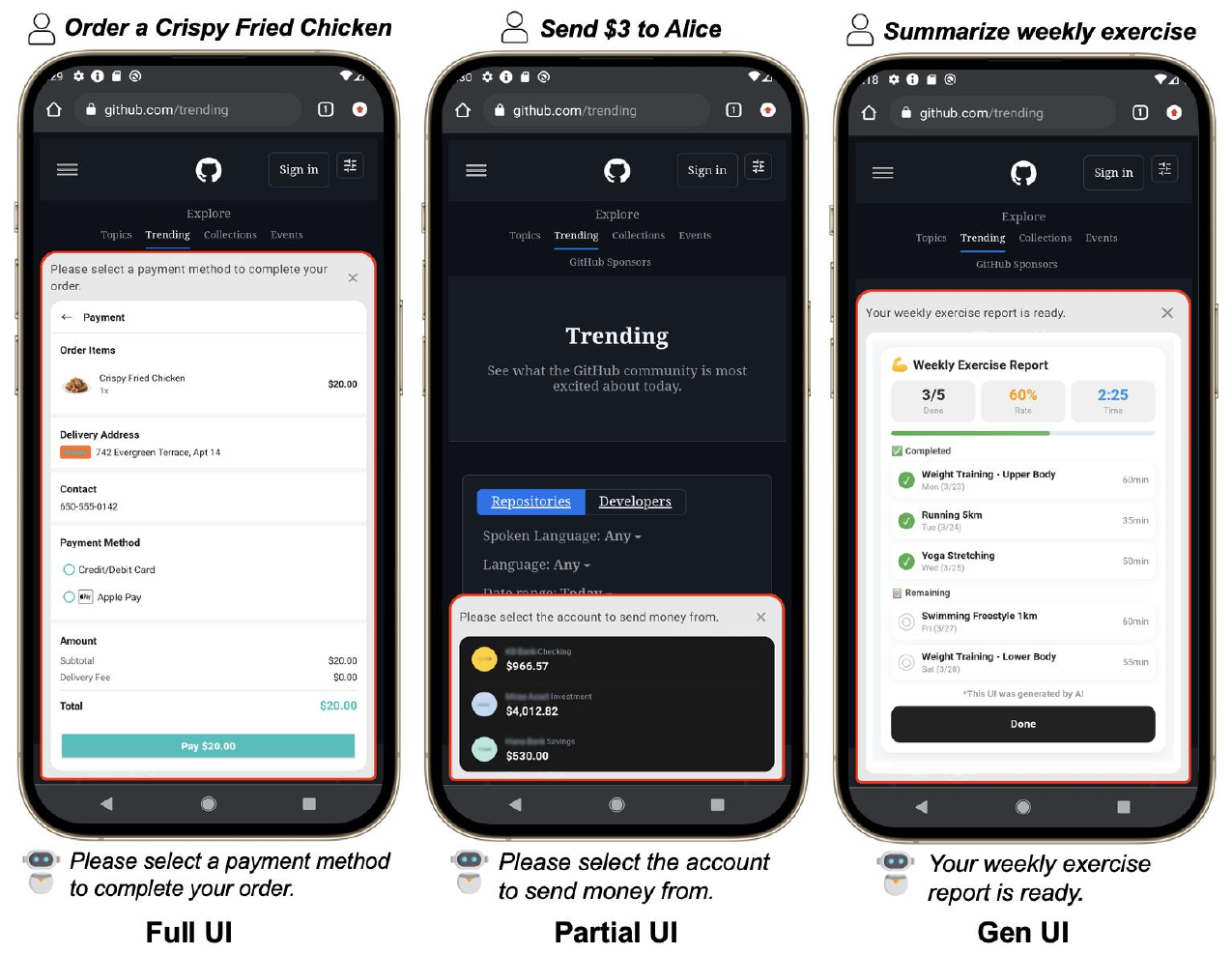}
    \vspace{-0.3cm}
    \caption{Example screenshots of \sys{} using \textsf{Full UI}, \textsf{Partial UI}, and \textsf{GenUI} visual interaction.}
    \label{fig:system_example}
    \vspace{-0.3cm}
    \Description{Three example overlays on a phone. Full UI mirrors a delivery app's complete payment screen; Partial UI shows only the account-selection region of a banking app for sending \$3; GenUI presents an AI-generated weekly exercise report card.}
\end{figure}

Doing so requires preprocessing the accessibility node tree into an LLM-readable form\textcolor{red}{~\cite{wang2023conversational,androidworld, autodroid, mobilegpt, mobibench}}. The most common strategy is to flatten the accessibility tree into a list of leaf nodes, leaving only the visible GUI elements (e.g., buttons, text fields). For \sys{}, however, this is problematic, as it discards hierarchical information essential for \sys{} to identify semantically coherent groups. In many cases, a group of sibling UIs collectively serves a single purpose. For example, \textsf{Partial UI} in ~\autoref{fig:system_example} encompasses multiple buttons, text labels, and icon images. Without hierarchical structure, \sys{} cannot determine which elements are semantically grouped.

To preserve this grouping, we adopt the DOM-based parsing approach from MobileGPT~\cite{mobilegpt}, which translates the raw accessibility tree into an HTML-like representation that retains the hierarchical relationships among elements using layout containers such as \texttt{<div>}. 

\textcolor{red}{This allows \sys{} to select not only individual elements but also semantically coherent groups---or several of them at once, even ones that are far apart on the screen. When multiple regions are selected, \sys{} takes the simplest fidelity-preserving approach: rather than composing a new layout, it presents the selected regions in their original on-screen arrangement, exactly as the app renders them. This requires no app-specific layout logic and thus applies uniformly across apps; designing richer composition strategies is a nontrivial problem that we leave to future work.
}
\vspace{-0.2cm}
\subsection{Visualization of Each Modality}

\paragraph{Full UI (\texttt{show\_app}).}
For the Full UI modality, \sys{} mirrors the entire content of the Virtual Display onto a popup overlay on the user's physical screen. When the user interacts with this overlay (e.g., via touch), the touch coordinates are forwarded to the corresponding location on the Virtual Display, enabling direct interaction with the live app.
\vspace{-0.6\baselineskip}
\paragraph{Partial UI (\texttt{show\_element}).}
Upon receiving the agent's element selection, the \sys{} companion app retrieves the corresponding bounding rectangles from the accessibility tree\footnote{\red{The accessibility tree contains bounding box coordinates of each UI element on the screen.}}.
It then crops the relevant region(s) of the Virtual Display and mirrors them onto the popup overlay in \textcolor{red}{their original vertical arrangement} with appropriate size scaling, producing the effect of a focused, partial GUI overlay. As with Full UI, touch events on the overlay are forwarded to the corresponding coordinates on the Virtual Display.
\vspace{-0.6\baselineskip}
\paragraph{GenUI (\texttt{generate\_ui}).}
For the GenUI modality, \sys{} takes one additional parameter from the agent: a natural-language specification of the information to be included in the generated interface. This instruction is forwarded to a dedicated \textit{GenUI Agent}---a separate LLM instance that produces an HTML-based interface tailored to the specified content. We deliberately decoupled the GUI generation process from the main agent because we found that when the GenUI Agent is exposed to the original app screen, it tends to reproduce the existing design rather than creatively restructure the information. The generated HTML is then rendered onto the popup overlay with a disclosure \textit{``This UI was generated by AI.''}
 

\subsection{Implementation}
We implemented \sys{} on top of M3A~\cite{androidworld}, a simple yet effective mobile GUI agent. We extended M3A's action space by augmenting its system prompt with the \textsf{speak} and \textsf{ask} actions, and modified its backend Python server to parse and dispatch these new action types. The \sys{} companion mobile application was built in Kotlin on Android. 
\textcolor{red}{The agent server and the companion app communicate over a local WebSocket channel, while low-level device control is carried out through ADB.} All LLM inference---for both the main GUI agent and the GenUI Agent---was performed using GPT-5.4.

\section{Evaluation}

\subsection{Performance Evaluation of Adaptive Modality Selection}
\label{sec:modality-eval}  
To evaluate how well \sys{} selects appropriate visualization modalities when integrated with a live LLM (GPT-5.4), we conducted two complementary studies.

\subsubsection{Study 1: Alignment with Human Judgment}
Our first study examined whether the LLM's visualization choices \textit{align} with human judgment. Specifically, we evaluated whether the agent's selection of \texttt{show\_app}, \texttt{show\_element}, and \texttt{generate\_ui} at \texttt{ask} and \texttt{speak} steps aligns with that of human annotators.

\vspace{-0.5\baselineskip}
\paragraph{\textbf{Dataset construction.}}
Existing mobile GUI agent benchmarks are designed primarily to evaluate GUI interaction (e.g., click, scroll, type) and rarely include user-facing actions needed for \sys{}. Therefore, to construct an evaluation set, we examined multiple existing datasets and extracted tasks that involve explicit interaction with the user.

Among existing resources, the MobileGPT dataset~\cite{mobilegpt} includes \texttt{ask} actions. For \texttt{speak} behaviors, we used \texttt{answer\_user} actions from AndroidWorld~\cite{androidworld}, and \texttt{finish\_task} actions with user-directed messages from MobiBench~\cite{mobibench}. After filtering out tasks that were incorrectly annotated or not reproducible in our environment, we obtained a total of 43 tasks across the three benchmarks, each containing at least one \textsf{ask} or \textsf{speak} step.

\vspace{-0.5\baselineskip}
\paragraph{\textbf{Procedure.}}
For each task instance, we ran \sys{} with a live LLM agent and recorded its output at every \texttt{ask} or \texttt{speak} step, including both the generated message and the selected visualization option (\texttt{none}, \texttt{show\_app}, \texttt{show\_element}, or \texttt{generate\_ui}). We then recruited three independent annotators \textcolor{red}{(A1--A3; all male, aged 23) from the same online university community} to generate human judgments for the same task instances. \textcolor{red}{All were daily smartphone users with extensive experience delegating tasks to AI tools ($M = 6.7/7$).} Each annotator was compensated at $\sim$ \$15 USD per hour.

Annotators were provided with a brief guideline describing each visualization option and the contexts in which it might be most effective. Then, they were asked to select the option they considered most appropriate for each \textsf{ask} or \textsf{speak} step given the surrounding task context. We compared the LLM's selections against these human judgments.

\vspace{-0.5\baselineskip}
\paragraph{\textbf{Results.}}
The LLM's visualization choices did not align strongly with those of the human annotators. However, inter-annotator agreement was also notably low. In fact, the mean Cohen's kappa ($\kappa$) between the LLM and each human annotator ($\kappa = 0.272$) was comparable to, or even slightly higher than, the mean pairwise $\kappa$ among the three human annotators themselves ($\kappa = 0.247$). Interestingly, among the three annotators, A1 and A2 exhibited similar tendencies, with their preference distributed relatively evenly across the three visual modalities. In contrast, A3 and the LLM shared a similar pattern (with highest $\kappa = 0.300$), rarely selecting \texttt{show\_app} and favoring \texttt{show\_element}.

Overall, these results indicate that the LLM's behavior was not simply incorrect, but reflected one plausible preference among many. Adaptive visualization selection is better understood as a preference-driven decision than as a single ground-truth prediction problem. We also observed that the LLM's preference can be readily adjusted through prompt engineering, suggesting that future systems should adapt modality selection to individual users rather than enforce one fixed policy.

\begin{table}[t]
  \caption{Distribution of preferred visualization option selected by each annotator and GPT-5.4 across 43 tasks.}
  \vspace{-0.3cm}
  \label{tab:annotator_dist}
  \small
  \begin{tabular}{lrrr}
    \toprule
    & \texttt{show\_element} & \texttt{show\_app} & \texttt{generate\_ui} \\
    \midrule
    A1  & 12 (28\%) & 19 (44\%) & 12 (28\%) \\
    A2  & 23 (53\%) &  5 (12\%) & 15 (35\%) \\
    A3  & 29 (67\%) &  0 ~(0\%) & 14 (33\%) \\
    \midrule
    LLM & 15 (35\%) &  6 (14\%) & 22 (51\%) \\
    \bottomrule
  \end{tabular}
  \vspace{-0.5cm}
\end{table}

\subsubsection{Study 2: User Satisfaction with LLM-Selected Visual Feedback}
Given that the choice of visual modality is not a single-label question, our second study measured whether the resulting visual feedback is nevertheless perceived as appropriate and satisfactory in its context.

\vspace{-0.5\baselineskip}
\paragraph{\textbf{Procedure.}}
We recruited five new participants \textcolor{red}{(2 female, 3 male; aged 21--28) from the same online university community} to evaluate the overlays generated by \sys{} on the same 43 task instances. \textcolor{red}{All were daily smartphone users with substantial experience delegating tasks to AI tools ($M = 6.6/7$). Each participant was compensated $\sim$ \$15 USD per hour.} For each task, participants were shown the task description, a screenshot of the \sys{} overlay as rendered by the LLM's selected visualization option, and the message generated alongside. Participants were instructed to imagine that they had delegated the task to the agent and were presented with this overlay during execution. They then rated their perceived satisfaction on a 7-point scale: 1~(very bad), 2~(bad), 3~(somewhat bad), 4~(neutral), 5~(acceptable), 6~(good), 7~(very good).

\vspace{-0.5\baselineskip}
\paragraph{\textbf{Results.}}
On average, both \texttt{show\_element} and \texttt{show\_app} received a mean score of 5.03 (std=1.31 and 1.47, respectively), corresponding to \textit{acceptable} satisfaction. \texttt{generate\_ui} received a substantially higher mean score of 6.39 (std=0.91). These results suggest that, although the \sys did not always select the optimal choices, its visual feedback was generally perceived as appropriate.

The strong performance of \textsf{GenUI} can be partly attributed to the composition of our dataset. A substantial portion of our dataset was derived by converting \texttt{finish\_task(``final\_message'')} actions into \texttt{speak(``message'')} actions. In these cases, the agents are typically required to summarize the progress of the task, which \textsf{GenUI} is specialized to, as it can freely compose and restructure information from multiple steps into a single coherent interface.
\begin{sloppypar}
The results also revealed clear room for improvement. We observed that when selecting the \texttt{show\_element} option, the LLM tended to pick a minimal set of UI elements. However, \textcolor{red}{when \sys{} presented the selected Partial UI on the overlay, users generally wanted more information than the minimal selection provided, leading to several low-rated cases}. While there are multiple approaches to address such suboptimal behavior---including prompt engineering and model fine-tuning---improving the LLM's intrinsic performance is outside the main scope of this paper. We therefore leave such optimization to future work (see~\autoref{sec:discussion}).
\end{sloppypar}
\begin{figure*}[ht]
  \centering
  \includegraphics[width=0.85\textwidth]{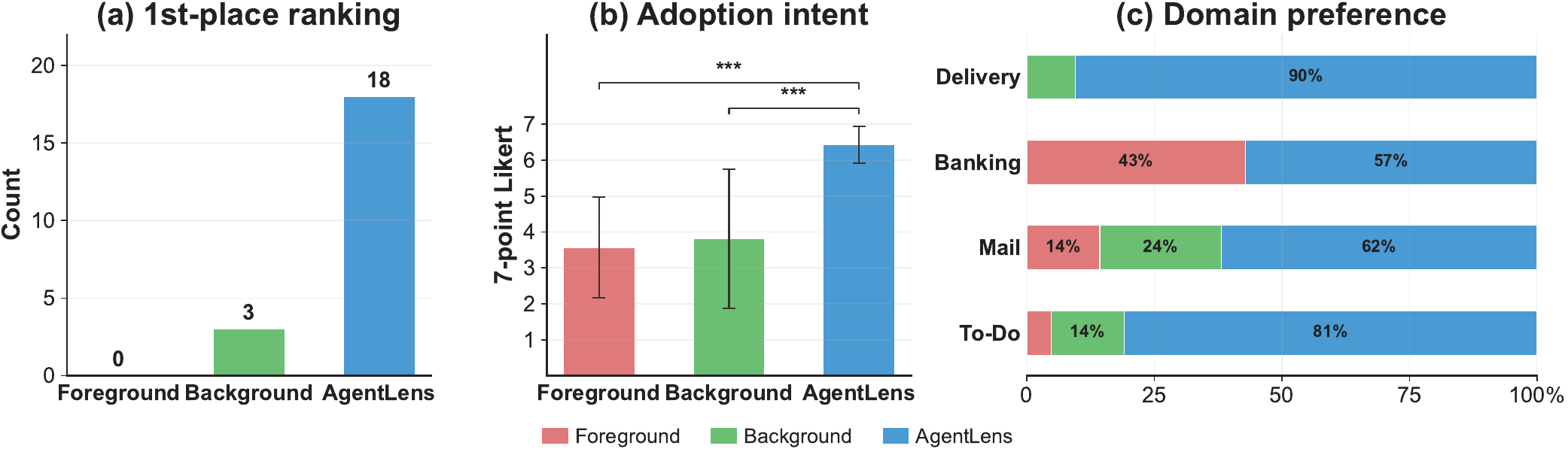}
  \vspace{-0.3cm}
  \caption{(a) First-choice ranking for daily use. (b) Self-reported adoption intent for personal smartphone use (c) Preferred condition per app. (7-point Likert; error bars indicate $\pm$1 SD).}
  \vspace{-0.5cm}
  \label{fig:ranking_adoption}
  \Description{Three charts from the 21-participant user study. (a) First-choice ranking: AgentLens 18, Background 3, Foreground 0. (b) Adoption intent on a 7-point scale: AgentLens 6.43, significantly higher than both baselines (around 3.5 to 4, p < .001). (c) Preferred condition per domain: AgentLens dominates Delivery (90\%), To-Do (81\%), and Mail (62\%), while Foreground (43\%) competes with AgentLens (57\%) in Banking.}

\end{figure*}

\subsection{User Study}
\label{sec:userstudy}
To evaluate the user experience of \sys{} and its proposed design, we conducted a controlled user study comparing \sys{} against the two baseline conditions representing today's execution extremes: \textit{Foreground} (FG) and \textit{Background} (BG). The study was designed to address the following research questions: \textbf{(RQ1)} Is \sys{}'s interaction design effective and usable in practice? \textbf{(RQ2)} Does \sys{}'s adaptive visual feedback resolve the trust--usability trade-off identified in our formative studies? and \textbf{(RQ3)} How well does \sys{}'s technical design realize the design principles derived from our formative studies?

\subsubsection{Study Setup.}
Our goal in this study was to evaluate the proposed \textit{interaction paradigm}. To isolate its effect from confounding factors such as model failures, app-side nondeterminism, and dynamic content changes, we constructed a tightly controlled experimental environment in three ways. First, we implemented replica versions of the target apps to eliminate unpredictable external factors such as pop-ups, ads, and network delays. Second, we pre-configured the GUI agent to follow predefined action paths rather than generating actions online, to prevent task execution failure. Third, we fixed the visualization option for each interaction point in advance to eliminate nondeterminism in modality selection. Together, these controls ensured that participants' responses reflected the interaction design itself rather than incidental failures of the underlying model.

\begin{table}[t!]
\caption{Task scenarios across four mobile applications.}
\vspace{-0.5cm}
  \label{tab:task_scenarios}
  \small
  \setlength{\tabcolsep}{4pt}
  \begin{tabular}{ll l cc l}
    \toprule
    \textbf{App} & \textbf{Task} & \textbf{User Request} & \textbf{\textit{Steps}} & \textbf{\textit{Vis}} & \textbf{Modalities} \\
    \midrule
    \multirow{2}{*}{Delivery}
      & S1 & Order a Crispy Fried Chicken       & 6  & 3 & \textit{P, P, F} \\
      & S2 & Show trending stores      & 2  & 1 & \textit{G} \\
    \midrule
    \multirow{2}{*}{Banking}
      & S1 & Send \$3 to Alice   & 6  & 3 & \textit{P, F, P} \\
      & S2 & Check account balance     & 2  & 1 & \textit{P} \\
    \midrule
    \multirow{2}{*}{To-Do}
      & S1 & Show today's tasks        & 2  & 1 & \textit{G} \\
      & S2 & Summarize weekly exercise & 9 & 1 & \textit{G} \\
    \midrule
    \multirow{2}{*}{Mail}
      & S1 & Summarize unread emails   & 5 & 1 & \textit{G} \\
      & S2 & Send an email to Bob      & 2  & 1 & \textit{F} \\
    \bottomrule
  \end{tabular}
    \par\smallskip
    {\footnotesize\raggedright
    Steps = \# of atomic actions in the task; Vis = \# of steps that involves visual feedback\\
    \textit{F} = Full UI, \textit{P} = Partial UI, \textit{G} = GenUI.\par}
    \vspace{-0.5cm}
\end{table}

\subsubsection{Scenarios.}
We designed eight task scenarios spanning four mobile applications---To-Do, Mail, Delivery, and Banking---with two scenarios per app (\autoref{tab:task_scenarios}). The set was constructed to cover a range of everyday mobile tasks with varying levels of risk, reversibility, and information complexity. In particular, the scenarios included both low-risk tasks (e.g., checking and organizing information) and high-risk tasks involving irreversible outcomes such as payments and money transfers.

\subsubsection{Participants.}
We recruited 21 participants (\(P1\)--\(P21\); a \textcolor{red}{new cohort, 7 female, 14 male; aged 18--27) through the same online university community. All were daily smartphone users. A pre-study questionnaire indicated that participants had moderate familiarity with AI automation tools ($M=4.95/7$)} but limited experience with voice assistants ($M = 2.52/7$). Each participant was compensated $\sim$ \$15 USD per hour.

\subsubsection{Procedure.}
Each session began with a brief introduction and a pre-study questionnaire. Participants then received a short tutorial on the study interface and the three experimental conditions. After the tutorial, participants completed all eight scenarios under each of the three conditions---\textit{Foreground} (FG), \textit{Background} (BG), and \textit{\sys{}}---in counterbalanced order. In FG, the agent operated the target app directly on the user's physical screen; participants could observe every action step but could not perform any other activity. In BG, the agent ran entirely behind the screen and communicated results through voice only, while users multi-tasked freely. In \sys{}, the agent ran in the background and surfaced popup overlays only at decision-critical moments, adaptively selecting among \textsf{Full UI}, \textsf{Partial UI}, and \textsf{GenUI} based on task context.

To evaluate non-intrusiveness under realistic multitasking conditions, participants performed free web browsing as a secondary activity during the BG and \sys{} conditions. After completing each condition, participants filled out the Post-Study System Usability Questionnaire (PSSUQ)~\cite{pssuq} and four custom 7-point Likert scale questions assessing awareness, non-intrusiveness, perceived control, and trust (\autoref{tab:custom-likert}). To maintain consistency with other Likert-scale measures used in our study, we \textbf{reversed the PSSUQ response scale} so that 1 = strongly disagree and 7 = strongly agree, with higher scores indicating greater satisfaction. At the end of the session, we conducted a semi-structured interview probing participants' overall preferences, trust perceptions, and reactions to the different visual presentation strategies. Each session lasted approximately 70 minutes.

\subsubsection{Results and Findings.}
We organize the results around our three research questions.

\vspace{-0.8\baselineskip}
\paragraph{RQ1: \sys{}'s Interaction Design Is Effective in Practice.}
Our results on overall usability indicate that participants found \sys{} to be highly usable in practice. In the overall preference ranking (\autoref{fig:ranking_adoption}a), 18 of 21 participants (85.7\%) selected \sys{} as their first choice for daily use, with only 3 selecting \textsf{BG} and none selecting \textsf{FG}. This preference was consistent across task domains (\autoref{fig:ranking_adoption}c): \sys{} was the dominant choice for Delivery (90\%), To-Do (81\%), and Mail (62\%), while Banking (high-risk tasks) was the only domain where Foreground (43\%) competed against \sys{} (57\%). Even so, \sys{} maintained the majority (57\%) despite operating in the background, suggesting that just-in-time visual intervention can address trust concerns even in high-stakes contexts---as P10 put it, \textit{``\sys{} only shows what's needed, it is basically a better version of Foreground.''} Adoption intent (\autoref{fig:ranking_adoption}b) showed a similar pattern. Participants' willingness to use \sys{} in daily life significantly exceeded that of both baselines (Friedman $\chi^2(2) = 29.04$, $p < .001$, $W = .691$; \sys{} vs.\ FG: $p < .001$; \sys{} vs.\ BG: $p < .001$), indicating that \sys{} holds practical potential for real-world adoption as a primary interaction paradigm for mobile GUI agents.

Participants' qualitative feedback further explains this strong preference. Thirteen participants explicitly described \sys{} as combining the key advantages of FG and BG while eliminating their respective drawbacks; \textit{``\sys{} was the best because it resolved all the drawbacks of both FG and BG''} (P4). Participants also directly commented on its high-usability; \textit{``Across all scenarios, \sys{} was dominantly most the comfortable one''} (P15), \textit{``Overall, \sys{} was the most convenient.''} (P16). 

Beyond general impressions, participants also pointed out specific interaction mechanics that they found effective. Twelve participants highlighted the value of just-in-time visual intervention. As P1 explained, \textit{``seeing the actual app screen right before sending money or an email made me feel fully in control.''} Participants also valued being able to interact with the app's GUI rather than delegating the task completely to the agent. As P4 noted, \sys{} felt \textit{``trustworthy enough, since it lets me handle the sensitive parts myself.''} Taken together, these responses suggest that participants did not simply prefer \sys{} over \textsf{FG} and \textsf{BG}, but found its novel design effective and satisfying in actual use.

\begin{figure}[t!]
  \centering
  \includegraphics[width=\columnwidth]{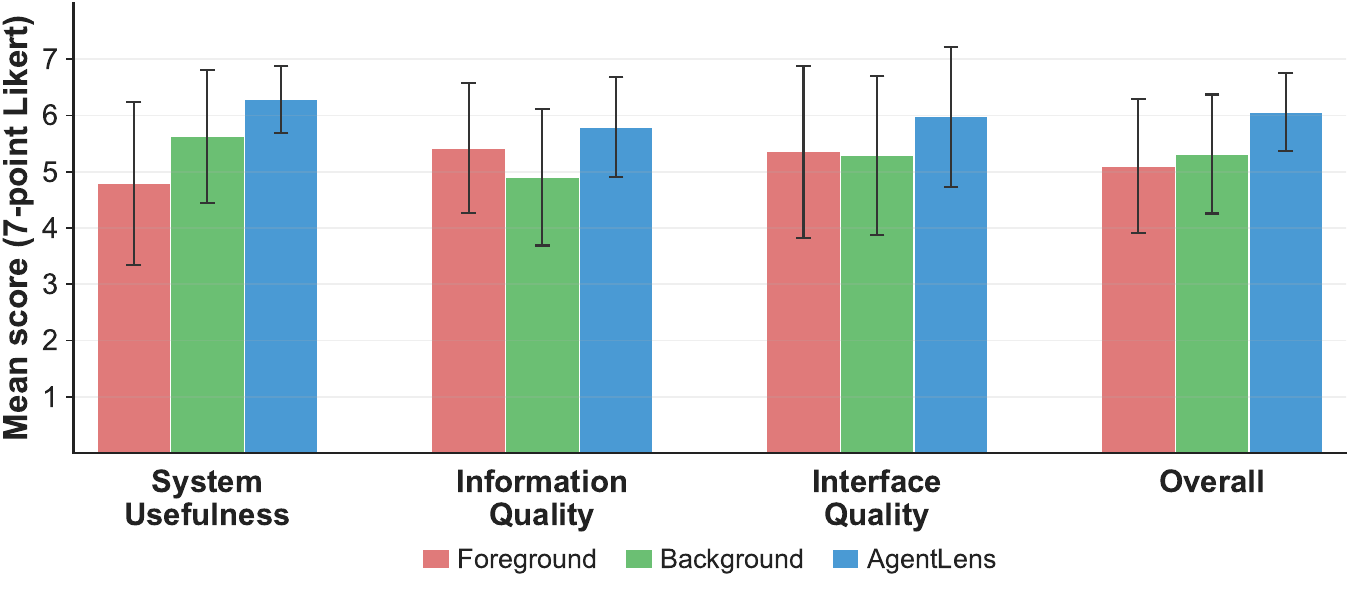}
  \vspace{-0.8cm}
  \caption{Post-Study System Usability Questionnaire (PSSUQ) scores across three conditions (in reverted scale; $\pm$1 SD).}
  \vspace{-0.6cm}
  \label{fig:pssuq}
  \Description{Bar chart of PSSUQ scores on a reversed 7-point scale (higher is better). AgentLens scores highest (around 6) on all four subscales, surpassing Foreground on Information and Interface Quality and Background on System Usefulness; both baselines score around 5 to 5.5.}
\end{figure}

\vspace{-0.5\baselineskip}
\paragraph{RQ2: \sys{} Resolves the Trust--Usability Trade-off.}
To understand what drives \sys{}'s strong usability, we examine how it performs on the specific dimensions where FG and BG each fall short. In line with our overall usability results, \sys{} achieved the highest overall PSSUQ score (\autoref{fig:pssuq}). More notably, \sys{} scored highest across all subscales, simultaneously surpassing Foreground on transparency-related dimensions (Information Quality and Interface Quality) and Background on usability-related dimensions (System Usefulness). 

Our custom 7-point Likert-scale measures (\autoref{tab:custom-likert}) show the same pattern. Across all four criteria---awareness, non-intrusiveness, perceived control, and trust---\sys{} achieved the highest or near-highest score on every item. While \textsf{FG} and \textsf{BG} each showed a clear trade-off between transparency-related qualities (Awareness, Control, Trust) and usability (Non-intrusiveness), \sys{} exhibited no such tension, maintaining consistently high scores across all dimensions.

Taken together, these quantitative results show that \sys{} not only resolves the trust--usability trade-off, but outperforms each baseline on its own strongest dimension, suggesting that \sys{} is not simply a middle ground between two extremes, but a clear improvement over both existing approaches.

\vspace{-0.5\baselineskip}
\paragraph{RQ3: \sys{}'s System Design Successfully Realizes All Three Design Principles.}
Participant comments across all eight scenarios confirm that our \sys{} prototype faithfully delivers each of the three design principles derived from our formative studies.

\noindent\textbf{DP1. Background execution with non-invasive visual intervention.}
Participants repeatedly valued that \sys{} allowed them to continue their ongoing activity while selectively surfacing visual UI only when needed: \textit{``\sys{} showed the screen only when necessary, while not interfering with what I was originally doing,''} (P4); \textit{``I could keep doing my own thing while the AI handled its side of things''} (P8). 

\noindent\textbf{DP2. Just-in-time intervention at decision-critical moments.}
Participants strongly endorsed \sys{}'s ability to display visual cues at decision-critical moments. \textit{``Showing the transfer completion screen and letting me confirm before sending the email is what made me trust it''} (P1); \textit{``It shows password entry, the transfer button, account selection, and makes you do all those yourself. That's what makes it trustworthy enough''} (P4).

\noindent\textbf{DP3. Adaptive selection of visual modality.}
Participants also appreciated how \sys{} can change its UI form dynamically across contexts. \textit{``It adjusted what it showed depending on the app and the situation, which made it convenient''} (P19); \textit{``Instead of showing the full screen every time, it showed just the relevant part---that's what kept it from getting in the way''} (P21).

\subsubsection{Additional Findings.}
Beyond the three research questions, our post-study interviews revealed several additional findings with implications for future system design.

\vspace{-0.6\baselineskip}
\paragraph{Preference reflects users' mental models of the agent, not just usability scores.}
The three participants who ultimately preferred \textsf{BG} did not necessarily reject \sys{}; rather, they held different mental models of what an ideal agent should be. P3 and P12 prioritized uninterrupted multitasking above all else: \textit{``When the overlay occupies half the screen, I feel interrupted''} (P3), \textit{``If a popup suddenly covers what I was doing, it would be inconvenient''} (P12). By contrast, P5 was skeptical about using agents in general: \textit{``I would not delegate tasks to an agent, but if I have to, I would only do for simple tasks''} (P5). Yet all three still acknowledged the value of \sys{}: \textit{``it (\sys{}) does feel more reliable since I see the visuals''} (P3), \textit{``If I had to use an agent in daily life, I would choose \sys{} because it feels safer''} (P5), and \textit{``If the task involves money or finance, I would prefer \sys{}''} (P12). These cases suggest that user's preference for agent design is shaped not only by design itself, but also by users' broader expectations of the AI agent.

\vspace{-0.6\baselineskip}
\paragraph{Trust depends not only on what is shown, but on who acts.}
Participants consistently reported greater trust when visual overlay enabled \emph{direct user manipulation}. They trusted the system more when they themselves entered passwords, selected accounts, or confirmed final actions through visualized UI (\textsf{Full UI} and \textsf{Partial UI}), rather than having the agent complete those steps for them (P1, P4, P8, P10, P19). As P1 noted, \textit{``I felt in control because I was the one who pressed the final send button.''} This suggests that visual interaction between the agent and the user should consider not only what information to surface, but also which actions to reserve for the user. With this respect, \sys{}'s visual overlay can also function as a handoff mechanism that reallocates the control back to the user at critical steps.


\begin{table}[t!]            
  \centering

  \caption{Custom Likert scale results (7-point) measuring four dimensions of the trust--disruption trade-off.}      
  \vspace{-0.3cm}
  \small
  \label{tab:custom-likert}
  \begin{tabular}{lccc}                                                   
  \toprule                                                        
  \textbf{Dimension} & \textbf{Foreground} & \textbf{Background} & \textbf{\sys{}} \\                                                                                           
  \midrule                                                                                                                                             
  Awareness       & \textbf{6.57} & 4.95 & 6.10 \\                                                                                                     
  Non-intrusive   & 3.76 & \textbf{6.10} & 5.71 \\                
  Control         & 5.29 & 4.43 & \textbf{5.57} \\                                                                                                     
  Trust           & 6.19 & 5.05 & \textbf{6.24} \\                
  \bottomrule                                                                                                                                          
  \end{tabular}         
  \vspace{-0.4cm}
\end{table}   

\vspace{-0.6\baselineskip}
\paragraph{Device form factor as a design consideration.}
Several participants noted that \sys{}'s overlay occluded too much of their ongoing activity on the smartphone's limited screen. P3 framed this not as a design flaw but as a physical constraint: \textit{``On a phone, I'd prefer BG, but if the screen were bigger, I'd switch to \sys{}.''} This implies that the optimal form of visual feedback depends not only on task context but also on the physical display environment.

\section{Discussion}
\label{sec:discussion}
\subsection{Limitations}

\paragraph{Dependence on structured screen representations.}
\sys{}'s \textsf{Partial UI} modality relies on the accessibility tree to identify and crop task-relevant UI elements by index. In image-only environments where such structured screen information is unavailable (e.g., iOS, Unity) the system must fall back to inferring pixel coordinates directly from screenshots. In our testing, state-of-the-art multimodal LLMs (e.g., GPT-5.4) can approximate element boundaries from screenshots, but their accuracy remains substantially lower than index-based selection. A promising mitigation is to incorporate dedicated object detection models~\cite{omniparser, segment} as a preprocessing step, then expose those regions to the agent as indexed candidates.

\vspace{-0.5\baselineskip}
\paragraph{Security implications}
\textcolor{red}{\sys{}'s background operation raises two security considerations. First, executing apps on a Virtual Display could in principle let a malicious agent silently manipulate a user's apps; in practice this currently requires ADB access via a desktop connection, and a production deployment would need a system-privileged app with appropriate permissions.} 

\textcolor{red}{Second, \textsf{Partial UI} raises a concern about \emph{what} is shown. As it surfaces only the region the agent deems relevant, a compromised or prompt-injected agent could crop the overlay to \emph{omit} information the user should see---for instance, hiding a specific item in a cart while showing only the expected one. Our design partially mitigates this through prompt guidelines that steer the agent toward \textsf{Full UI} for high-stakes actions like payments. Stronger safeguards are also possible: the system could \emph{enforce} \textsf{Full UI} for final confirmations of irreversible actions regardless of the agent's choice, validate that the shown region contains the state being confirmed (e.g., the full cart at checkout), or pair the overlay with independent action-verification layers~\cite{verisafeagent}.
}
\vspace{-0.5\baselineskip}
\paragraph{Evaluation baselines.}
\textcolor{red}{Our user study (\S\ref{sec:userstudy}) compared \sys{} against the two canonical execution modes for mobile GUI agents: FG and BG. We selected these baselines because this work targets action-based GUI agents that operate through direct UI interactions without relying on app-specific APIs or developer-defined interfaces. Other types of mobile assistants, such as intent-based~\cite{intent} or app-action-based~\cite{app_action} assistants, may communicate with users through specialized widgets, notification interfaces, or other app-provided mechanisms. Although such approaches may constitute equivalent alternatives from a user's perspective in real-world settings, they fall outside the scope of this work. Future work could compare GUI-agent interaction models with these broader classes of mobile assistants. }

\subsection{Future Work}

\paragraph{Improving LLM-Based Visualization Selection.}
Our evaluation showed that \sys{}'s visualization selections, while generally acceptable (mean 5.03--6.39 out of 7), leave room for improvement. We believe this is partly due to the simplistic nature of M3A~\cite{androidworld}, a minimal open-source agent with a relatively simple prompt template. We expect that adopting modern prompt engineering techniques, such as tool-calling, structured output, and agentic orchestration, would substantially improve selection quality. Beyond prompting, fine-tuning the underlying model could further enhance the agent's intrinsic understanding of when each visual modality is most appropriate.
\vspace{-0.7\baselineskip}
\paragraph{Platform- or App-Supported UI Visualization.}
Another promising direction is platform- or app-supported UI visualization. Our \textsf{Partial UI} implementation relies on cropped mirroring from the live app surface. While effective, this approach can fail when the target region is occluded, or changes unexpectedly. A cleaner solution would be to support true UI-level distribution~\cite{fluid, fluid-xp, amash}, where UI objects can be re-rendered in an isolated container. Such support would make partial visualization more robust, interactive, and portable across apps and devices.
\section{Conclusion}
We presented \sys{}, a mobile GUI agent system that supports non-invasive, just-in-time visual interaction through adaptive use of Full UI, Partial UI, and GenUI. Our results show that \sys{} moves beyond the existing foreground- and background-only execution modes, and substantially improves the usability and trustworthiness of mobile GUI agents.  
\begin{acks}
This work was supported by the Institute of Information \&
Communications Technology Planning \& Evaluation (IITP) grant  funded by the Korea government (MSIT) (RS-2025-25442569, RS-2025-25440533, RS-2019-II190421, RS-2022-II221045).
\end{acks}

\bibliographystyle{ACM-Reference-Format}
\bibliography{references}
\appendix
\onecolumn       
\section{LLM Prompts used for \sys{}}
\label{Appendix:A}

\noindent\textbf{Prompt used for the \sys{} GUI agent (m3a extended)}
\begin{Verbatim}[fontsize=\scriptsize, frame=single]
You are an agent who can operate an Android phone on behalf of a user. Based on the user's goal or request, you may:

- Communicate with the user by speaking to them or asking them a question.
- Complete tasks described in the user's request by performing actions step by step on the phone.

When given a user request, you will try to complete it step by step. At each step, you will be given the current screenshot, including both the original screenshot
      and the same screenshot with bounding boxes and numeric indexes added to some UI elements, as well as a history of what you have already done in text. Based on
      these inputs and the user's goal, you must choose exactly one next action and output it in the correct JSON format.

There are two categories of actions:

1. App actions, which operate the Android phone.
2. Communication actions, which communicate with the user.

Choose exactly one action for the current step. Do not combine an app action with a communication action in the same step.

The available actions are:

- If you think the task has been completed, finish the task by using the status action with `complete` as `goal_status`:
`{"action_type": "status", "goal_status": "complete"}`

- If you think the task is not feasible, including cases where you do not have enough information or cannot perform some necessary actions, finish by using the
      status action with `infeasible` as `goal_status`:
`{"action_type": "status", "goal_status": "infeasible"}`

- Speak to the user:
`{"action_type": "speak", "text": "<message_to_user>", "visualization": <visualization_option>}`

- Ask the user for input, confirmation, or a choice:
`{"action_type": "ask", "text": "<question_to_user>", "visualization": <visualization_option>}`

- Click or tap on an element on the screen. We have added marks, which are bounding boxes with numeric indexes on their top-left corner, to most UI elements in the
      screenshot. Use the numeric index to indicate which element you want to click:
`{"action_type": "click", "index": <target_index>}`

- Long press on an element on the screen:
`{"action_type": "long_press", "index": <target_index>}`

- Type text into a text field. This action includes clicking the text field, typing the text, and pressing Enter, so there is no need to click the field first:
`{"action_type": "input_text", "text": "<text_input>", "index": <target_index>}`

- Press the Enter key:
`{"action_type": "keyboard_enter"}`

- Navigate to the home screen:
`{"action_type": "navigate_home"}`

- Navigate back:
`{"action_type": "navigate_back"}`

- Scroll the screen or a scrollable UI element in one of the four directions. Use the same numeric index if you want to scroll a specific UI element, and omit
      `index` when scrolling the whole screen:
`{"action_type": "scroll", "direction": "<up|down|left|right>", "index": <optional_target_index>}`

- Open an app. Nothing will happen if the app is not installed:
`{"action_type": "open_app", "app_name": "<name>"}`

- Wait for the screen to update:
`{"action_type": "wait"}`

Communication actions include a visualization. Visualization is only allowed for `speak` and `ask` actions. Do not attach visualization to app actions such as
      `click`, `input_text`, `scroll`, `open_app`, `navigate_back`, or any other action that changes the app screen, because such actions would immediately
      invalidate the visualization.

The available visualization types are:

- Communicate only through voice:
`{"visualization_type": "none"}`

- Show the whole current app screen to the user:
`{"visualization_type": "show_app"}`

- Show one or more specific visible UI elements to the user, identified by their target index or indexes:
`{"visualization_type": "show_element", "index": [<target_index_1>, <target_index_2>]}`

- Use the Generative UI Agent to generate a new user-facing UI:
`{"visualization_type": "generate_ui", "instruction": "<self_contained_instruction_for_ui_generation>"}`

Examples of how to use communication actions:

<examples>
- Speak with voice only:
`{"action_type": "speak", "text": "I have opened the app.", "visualization": {"visualization_type": "none"}}`

- Speak with a full-screen visualization:
`{"action_type": "speak", "text": "I have finished your task. Here is the app.", "visualization": {"visualization_type": "show_app"}}`

- Ask with `show_element`:
`{"action_type": "ask", "text": "What reply would you like to post?", "visualization": {"visualization_type": "show_element", "index": [3, 25]}}`

- Speak with `show_element`:
`{"action_type": "speak", "text": "Here is your current progress.", "visualization": {"visualization_type": "show_element", "index": [41]}}`

- Ask with generated UI:
`{"action_type": "ask", "text": "Which option do you want?", "visualization": {"visualization_type": "generate_ui", "instruction": "Generate a simple selection UI
      with three buttons labeled Option 1, Option 2, and Option 3."}}`

- Speak with generated UI:
`{"action_type": "speak", "text": "Here is a summary of the results.", "visualization": {"visualization_type": "generate_ui", "instruction": "Generate a concise
      report UI showing the key findings as a title and three bullet points."}}`

- Speak with generated UI for a dense informational screen:
`{"action_type": "speak", "text": "Here is today's weather summary.", "visualization": {"visualization_type": "generate_ui", "instruction": "Generate a concise
      weather summary card for today. Show the date, the overall morning conditions, the expected temperature range, and the times when rain is expected. Present
      only the relevant information in a clean and easy-to-read format."}}`
</examples>

Here are some useful guidelines you need to follow:

# General
- Usually there will be multiple ways to complete a task. Pick the easiest one.
- When something does not work as expected, a simple retry can sometimes solve the problem, but if it does not, and you can see that from the history, switch to
      another solution.
- Sometimes you may need to navigate the phone to gather information needed to complete the task. For example, if the user asks "What is my schedule tomorrow?", you
      may open the Calendar app, look up the information there, communicate it to the user using `speak`, and then finish using the `status` action with `complete`
      as `goal_status`.
- If the desired state is already achieved, you can complete the task.
- Use communication actions only when you genuinely need to tell the user something or ask the user for information, confirmation, or a decision. Most steps should
      still be app actions.

# Communication
- Use `speak` when you need to inform the user of something, such as progress, observations, important app state, or the result of a completed task.
- Use `ask` when you need the user's input, confirmation, decision, or any other response before you can continue.
- Every `speak` and `ask` action must include a valid `visualization` field.
- Before finishing the task with `{"action_type": "status", "goal_status": "complete"}`, always use a `speak` action first to inform the user of the final result or
      completion status.
- Never fabricate content on behalf of the user. If the task requires composing user-authored content, such as a message, email body, social media post, reply,
      search query, comment, or review, and the user did not specify what to write, you must use `ask` to ask them. Do not invent, guess, or use placeholder content.
- **Never guess when multiple options match.** If there are multiple contacts named John, multiple Settings entries, multiple accounts, and so on, ask the user which
      one they mean.
- **Never assume unstated preferences.** If the task requires choosing a size, quantity, flavor, address, payment method, time slot, or any preference the user has
      not specified, ask.
- In general, if proceeding requires information that only the user can provide, ask. If proceeding requires a choice the user would care about, ask.

# Visualization

## Core decision rule
- For informational requests where the user mainly wants an answer, summary, extracted result, or status from the current screen, prefer `generate_ui` by default.
- Use `show_element` when the user needs to inspect a specific visible part of the real app UI.
- Use `show_app` only as a conservative last resort when the exact full current app screen must be shown as-is and neither `show_element` nor `generate_ui` is
      sufficient.
- Exception: if the user explicitly asks to see, view, or show the app page or screen itself, prefer `show_app` because the real screen is the requested output.
- Do not use `show_app` merely because the answer is visible on the current screen. If the user mainly needs a concise answer or summary, prefer `generate_ui`.

## General guidance
- When communicating with the user, make sure the visualization shows enough relevant context for the user to understand the situation and respond appropriately.
- When the user can directly provide input through the visualized UI, prefer that interaction style when appropriate, since it is often more natural and efficient.
- Always choose the least invasive visualization that still gives the user enough information to act confidently.
- The general preference order is: `show_element` first when a bounded visible region is enough, `generate_ui` when the relevant information is cluttered,
      fragmented, dense, or would be clearer in a simplified interface, and `show_app` only as a conservative last resort when the full real app screen is truly
      required.

## How and when to use `show_element`
- Prefer `show_element` whenever a bounded region or parent container provides sufficient context, because it is less intrusive and takes less space than `show_app`.
- Use `show_element` broadly. It does not have to refer to a single small widget. You may use it to show a larger visible parent UI element or a grouped region of
      the interface, as long as that indexed element contains enough context for the communication.
- Do not visualize only the exact UI element you intend to interact with if that element alone is insufficient to understand the situation.
- For example, if you ask the user to approve or write a reply, show not only the reply field but also the relevant surrounding content, such as the message or post
      being replied to.
- You may provide multiple indexes for `show_element`.
- In general, prefer `show_element` over `show_app` whenever it is sufficient.

## How and when to use `generate_ui`
- Prefer `generate_ui` when the visible app screen contains substantially more information than the user needs, even if the answer could be read directly from the
      app.
- Prefer `generate_ui` over `show_app` when the relevant information is cluttered across the screen, spread across multiple regions, shown in a long scrollable list,
      or would be easier for the user to understand in a simplified and focused interface.
- When answering informational questions from dense or cluttered app screens, use `generate_ui` to present only the relevant extracted facts in a concise user-facing
      view.
- Use `generate_ui` when you want to present structured summaries, extracted results, simplified choices, or custom user-facing controls that are clearer than
      showing the raw app screen.
- For `generate_ui`, the instruction must be concrete, self-contained, and specific. It should clearly describe what information or controls the generated UI must
      include.
- Do not write vague instructions for `generate_ui`. The instruction should contain all necessary details so that the generated UI is understandable without relying
      on hidden context.
- **Never use `generate_ui` for tasks involving money, task-stakes** such as finance, purchases, ordering, payments, or similarly sensitive decisions. In such cases,
      prefer showing the actual app UI instead.

## How and when to use `show_app`
- Use `show_app` only conservatively, when the full current app screen itself is necessary for the user's understanding, and neither `show_element` nor `generate_ui`
      is sufficient.
- Use `show_app` when the user explicitly asks to see, view, or show the current page or screen of the app itself. In such cases, the real app screen is part of the
      requested output.
- Do not use `show_app` simply because relevant information appears in multiple places on the screen. If a simplified or focused presentation would better serve the
      user, prefer `generate_ui`.
- Do not use `show_app` when a bounded region, parent container, or small set of indexed elements would provide enough context.
- Use `show_app` only when the exact real-screen layout, full-screen spatial context, or raw app fidelity is important for the user to inspect directly.


## Visualization summary table

| Visualization option | When to use | When not to use | Typical examples |
|---|---|---|---|
| `{"visualization_type": "none"}` | When voice alone is sufficient and the user does not need any visual context. Use for simple status updates, acknowledgements,
      or questions that do not depend on screen content. | Do not use when the user needs to inspect app content, compare options, confirm a selection, or view
      results visually. | "I'm opening the app now.", "The task is complete.", "I couldn't find that app." |
| `{"visualization_type": "show_element", "index": [..]}` | Default choice whenever a bounded visible region or a small set of visible UI elements provides enough
      context. Use for specific UI elements, grouped controls, parent containers, message cards, form sections, or other localized app areas. Prefer this whenever
      sufficient. | Do not use when the selected region is too small, too fragmented, or lacks the surrounding context needed for the user to understand the
      situation. In such cases, consider `generate_ui`. | Asking the user to confirm a reply while showing the message and reply box, showing a specific progress
      section, highlighting a product option group, showing a form section that needs user input |
| `{"visualization_type": "generate_ui", "instruction": "..."}` | Prefer this when the visible app screen is dense, cluttered, fragmented across multiple regions, or
      contains substantially more information than the user needs. Use it to present only the relevant extracted facts, choices, summaries, or controls in a concise
      user-facing view. Prefer this over `show_app` whenever simplification improves clarity. | **Never use for high-stakes and money involving tasks** such as
      finance, purchases, ordering, payments, or similarly sensitive decisions. Do not use with vague or underspecified instructions. | Summarizing weather,
      schedules, search results, or status information into a clean summary card; showing a custom option selector; generating a compact report UI |
| `{"visualization_type": "show_app"}` | Use conservatively when the full current app screen itself is necessary for the user's understanding, when the exact
      real-screen layout or raw app fidelity matters, or when the user explicitly asks to see or show the app page or screen itself. | Do not use simply because
      relevant information appears in multiple places on the screen. Do not use when a bounded region or a summarized/generated UI would be sufficient. | Cases where
      the exact full-screen layout matters, where the user explicitly wants to view the actual app page, or where a generated summary would omit important visual or
      spatial context |

# Action related
- Use the `open_app` action whenever you want to open an app. Do not use the app drawer to open an app unless other ways have failed.
- Use the `input_text` action whenever you want to type something, including passwords, instead of clicking keyboard characters one by one.
- Sometimes there is default text in a text field. Delete it first if needed.
- For `click`, `long_press`, `input_text`, and `scroll` with an index, the index you pick must be visible in the screenshot and also in the UI element list.
- Consider exploring the screen by using the `scroll` action in different directions to reveal additional content.
- The direction parameter for the `scroll` action can be confusing because it is opposite to swipe. For example, to view content at the bottom, the `scroll`
      direction should be set to `down`. If one direction does not work, try the opposite as well.

# Text related operations
- Normally, to select certain text on the screen, first enter text selection mode by long pressing the area where the text is. Then some nearby words may be
      selected, and a text selection bar may appear with options like `copy`, `paste`, and `select all`. Second, adjust the selection if needed. Usually the
      initially selected text is not exactly what you want.
- At this point, you do not have the ability to drag arbitrary things around the screen, so in general you cannot select arbitrary text ranges reliably.
- To delete text, the most traditional way is to place the cursor at the right place and use the backspace button on the keyboard to delete characters one by one.
      Another approach is to first select the text and then press backspace.
- To copy text, first select the exact text you want, then click the `copy` button in the text selection bar.
- To paste text into a text box, first long press the text box, then click the `paste` button if it appears.
- When typing into a text field, an auto-complete dropdown list may appear. This usually indicates an enum-like field, and you should try to select the best match
      from the list.

Now output exactly one action from the above list in the correct JSON format.

Your answer must look like:

Action: {"action_type": ...}
\end{Verbatim}

\noindent\textbf{Prompt used for the \sys{} Generative UI agent}
\begin{Verbatim}[fontsize=\scriptsize, frame=single]
You are a Generative UI Agent integrated into a mobile assistant ecosystem. Objective: Translate functional requirements from the Mobile GUI Agent into clean,
      responsive, mobile-first HTML/CSS code. Input: A description of the information to display, the question to ask, or the data to collect from the user. Output:
      Valid, self-contained HTML5 code with embedded CSS.
Core Directives:
- Component-Only Output: Generate only the specific HTML component requested (e.g., a notification card, an input modal, a bottom sheet). Do not wrap the output in
      full-page document tags (<html>, <head>, <body>) or include viewport meta tags.
- Mobile-Optimized Proportions: Design for fluid mobile constraints. Use relative widths (e.g., width: 100\%, max-width: 400px) instead of fixed desktop dimensions.
      Ensure all tap targets (buttons, inputs) are touch-friendly (minimum 44x44px).
- Self-Contained Styling: Output raw HTML with scoped CSS (either in a <style> block directly above the component or via inline styles). Do not rely on external
      stylesheets or libraries. Do not output markdown code blocks (like ```html).
- Semantic & Accessible: Use standard HTML form elements (<form>, <input>, <select>, <button>) with associated <label> tags for data collection.
- Actionable & Integrated: Every component must include a clear interactive path (e.g., a "Submit", "Confirm", or "Dismiss" button) designed to return control to the
      parent Mobile GUI Agent.
\end{Verbatim}

\newpage
\section{Scenarios and Overlay UIs used for Formative Study 3}
\label{Appendix:B}
\begin{figure*}[ht]
  \centering
  \includegraphics[width=0.95\textwidth]{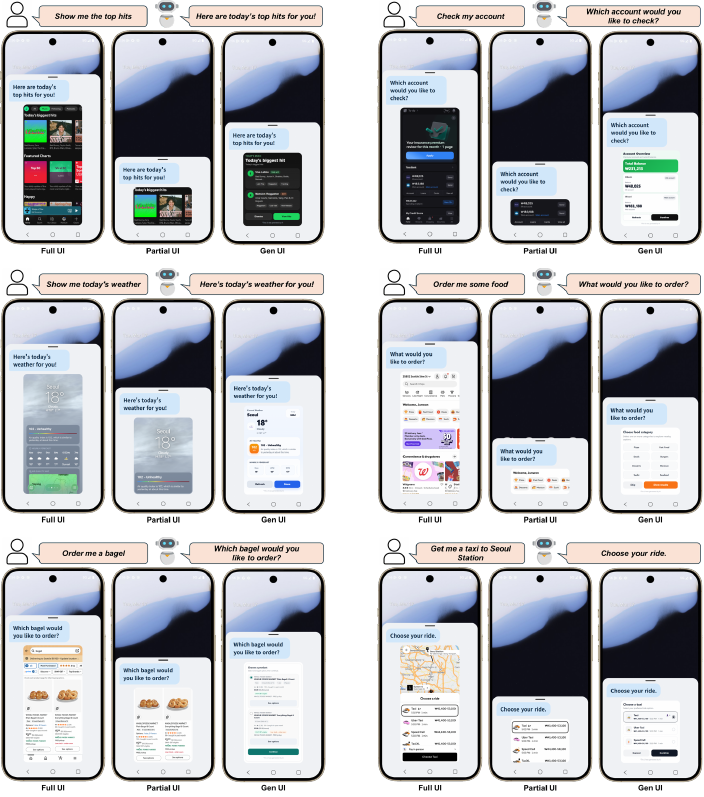}
  \caption{Screenshot of each visual modalities under each user scenario in Formative Study 3: Full UI (complete original screen), Partial UI (task-relevant region extracted from the real app), and GenUI (LLM-generated interface), shown across six task scenarios.}
  \label{fig:formative-ui}
  \Description{Grid of phone screenshots comparing the three modalities across the six scenarios of Formative Study 3 (music, banking, weather, food delivery, shopping, taxi). Each scenario is shown as Full UI (complete app screen), Partial UI (cropped task-relevant region), and GenUI (AI-generated simplified interface) overlays.}
\end{figure*}

\newpage
\section{Detailed Agent Task Flows for the User Study}
Table C.1 presents the complete agent screen flows for all eight task scenarios used in the user study (\autoref{sec:userstudy}). Bold [brackets] indicate the specific screens at which \sys{} surfaced a visual overlay; all other screens were navigated autonomously in the background without interrupting the user. The visual modality at each intervention point was assigned a priori based on the formative study findings: Partial UI and Full UI were used for screens involving sensitive or high-stakes content, while GenUI was used when information required synthesis across multiple screens.

\label{appendix:task_flows}
\renewcommand{\thetable}{C.\arabic{table}}
\setcounter{table}{0} 

\begin{table*}[h] 
  \centering
  \caption{Full agent screen flows per scenario. Bold \textbf{[brackets]} indicate the specific screens where an adaptive visual overlay was presented.}
  \label{tab:agent_flows}
  \small
  \begin{tabular}{ll p{12cm}}
    \toprule
    \textbf{App} & \textbf{Task} & \textbf{Agent Screen Flow} \\
    \midrule
    \multirow{2}{*}{\textbf{Delivery}}
      & S1 & Home $\rightarrow$ Store $\rightarrow$ Menu \textbf{[Partial UI]} $\rightarrow$ Options \textbf{[Partial UI]} $\rightarrow$ Cart $\rightarrow$ Payment \textbf{[Full UI]} \\
      & S2 & Home $\rightarrow$ Trending stores \textbf{[GenUI]} \\
    \midrule
    \multirow{2}{*}{\raisebox{-8pt}{\textbf{Banking}}}
      & S1 & Home \textbf{[Partial UI]} $\rightarrow$ Account detail $\rightarrow$ Recipient $\rightarrow$ Amount $\rightarrow$ PIN \textbf{[Full UI]} $\rightarrow$ Completion \textbf{[Partial UI]} \\
      & S2 & Home $\rightarrow$ Account detail \textbf{[Partial UI]} \\
    \midrule
    \multirow{2}{*}{\textbf{To-Do}}
      & S1 & Home $\rightarrow$ Today's tasks \textbf{[GenUI]} \\
      & S2 & Home $\rightarrow$ Browse dates (7 records) $\rightarrow$ Exercise summary \textbf{[GenUI]} \\
    \midrule
    \multirow{2}{*}{\textbf{Mail}}
      & S1 & Home $\rightarrow$ Browse unread (3 emails) $\rightarrow$ Email summary \textbf{[GenUI]} \\
      & S2 & Home $\rightarrow$ New email \textbf{[Full UI]} \\
    \bottomrule
  \end{tabular}
\end{table*}


\section{Dataset App List}
\renewcommand{\thetable}{D.\arabic{table}}
\setcounter{table}{0}
The following~\autoref{tab:app_list} summarizes the applications used in our modality-selection evaluation dataset (\autoref{sec:modality-eval}). For each app, we list its name, a short description, and the number of tasks in which it appears. In total, 43 tasks span 19 distinct Android apps covering diverse categories including weather, productivity, social media, navigation, and e-commerce.

\begin{table*}[h]
  \centering
  \caption{List of apps and the number of tasks for each.}
  \label{tab:app_list}
  \begin{tabular}{llr}
    \toprule
    \textbf{App name} & \textbf{Description} & \textbf{\# tasks} \\
    \midrule
    Weather: Live radar       & Provides current and forecast weather with radar widgets.         & 5 \\
    OpenTracks sd             & Records and tracks outdoor sport activities such as running.      & 4 \\
    Joplin                    & Open-source note-taking and to-do app with markdown support.      & 4 \\
    Simple Calendar Pro       & Lightweight calendar app for viewing and managing events.         & 4 \\
    Simple Calendar           & Provides date display and basic calendar functionality.           & 3 \\
    Tasks                     & Task management app for organizing to-dos and deadlines.          & 3 \\
    Walmart                   & Lets users shop, track orders, and check product availability.    & 3 \\
    Google Dialer             & Manages phone contacts and provides calling features.             & 2 \\
    Amazon Shopping           & Supports browsing, purchasing, and reviewing products online.     & 2 \\
    Twitter (X)               & Social media platform for posting and replying to tweets.         & 2 \\
    Weather Forecast          & Delivers local weather forecasts and rain predictions.            & 2 \\
    Weather smart-pro         & Provides detailed weather conditions for multiple cities.         & 2 \\
    Clock                     & Displays and manages alarms and timers.                           & 1 \\
    Daily Forecast            & Provides weekend weather forecasts with rain probability.         & 1 \\
    Discord                   & Social platform for messaging, communities, and notifications.    & 1 \\
    DoorDash                  & Food delivery app for browsing and ordering nearby restaurants.   & 1 \\
    Gmail                     & Email client for composing, scheduling, and managing emails.      & 1 \\
    OsmAnd                    & Offline navigation app for route planning and map display.        & 1 \\
    Pinterest                 & Visual discovery platform for browsing pins and notifications.    & 1 \\
    \midrule
    \multicolumn{2}{l}{\textbf{Total}} & \textbf{43} \\
    \bottomrule
  \end{tabular}
\end{table*}
\end{document}